\documentstyle[aps,prl,twocolumn,epsf,floats]{revtex}
\draft
\begin{document}
\wideabs{
\title{
Charged mobile complexes in magnetic fields: \\
A novel selection rule for magneto-optical transitions
}
\author{A. B. Dzyubenko}
\address{
 Institut f\"{u}r Theoretische Physik, J.W. Goethe-Universit\"{a}t,
 60054 Frankfurt,  Germany \\
General Physics Institute, Russian Academy of Sciences, Moscow 117942, Russia }
\author{A. Yu. Sivachenko}
\address{The Weizmann Institute of Science, Rehovot 76100, Israel
}
\date{JETP Lett. {\bf 70}, 514 (1999)}
\maketitle
\begin{abstract}
The implications of magnetic translations
for internal optical transitions of charged mobile electron-hole ($e$--$h$)
complexes and ions in a uniform magnetic field $B$ are discussed.
It is shown that transitions of such complexes are governed by a novel
exact selection rule.
Internal intraband transitions of two-dimensional (2D) charged excitons
$X^-$ in strong magnetic fields are considered as an
illustrative example.
\end{abstract}
}
 Recently, there has been considerable experimental and theoretical
interest in the behavior of 2D semiconductor complexes
--- negatively  $X^-$ ($2e$--$h$) and positively $X^+$  ($e$--$2h$)
charged excitons in magnetic fields
(see, e.g., Refs.~\cite{Exp,AHM,Whit97,EP2DS,P99} and references therein).
These three-particle bound states can be considered as analogs of the
hydrogen atomic $H^-$ and molecular $H_2^+$ ions, respectively.
The application of a magnetic field $B$ changes the hydrogenic
spectra drastically (see, e.g., Refs.~\cite{G&D,Simon,L&L80}).
As an example, whereas at $B=0$ the $H^-$ supports in 3D
only one bound singlet state, at any finite $B$ there appear \cite{Simon}
bound triplet $H^-$ states  and infinite number of quasi-bound states
(resonances) associated with higher Landau levels (LL's).
In 2D systems, it is the $X^-$ triplet
that becomes the ground state in high magnetic fields \cite{AHM,Whit97}.
Singlet and triplet $X^-$ and $X^+$ states are observed
in 2D systems by means of interband optical magneto-spectroscopy \cite{Exp}.
Intraband magneto-spectroscopy, in which internal transitions from
populated (ground) to excited states are induced by a photon,
can provide additional information about binding of hydrogen-like complexes.
In the $\sigma^+$ polarization internal transitions
in high $B$ are predominantly induced to the next electron LL.
Such photoionizing bound-to-continuum $X^-$ singlet and triplet transitions
in the far-infrared (FIR) have been predicted theoretically and recently
observed experimentally in GaAs quantum wells \cite{EP2DS}.
In this work we describe some general implications of
the existing exact symmetry --- magnetic translations ---
for internal transitions of charged mobile $e$--$h$ complexes in $B$.
We also present theory predictions for internal transitions
from the $X^-$ ground triplet state in the $\sigma^-$ polarization
in high $B$. A preliminary account of some of these results
and implications for interband magneto-optics of $X^-$ have been
reported in Ref.~\cite{P99}.

  Consider the Hamiltonian of interacting particles of charges $e_i$
in a magnetic field ${\bf B}$
\begin{equation}
                \label{H} 
   H = \sum_i \frac{ \hat{ \bbox{ \pi }}_i^2}{2m_i}
          + \case{1}{2} \sum_{i \ne j} U_{i j}({\bf r}_i-{\bf r}_j) \, ,
\end{equation}
here
$\hat{\bbox{\pi}}_i = -i\hbar \bbox{\nabla }_i -
\frac{e_i}{c} {\bf A}({\bf r}_i)$
and interparticle interactions potentials $U_{ij}$
can be rather arbitrary.
In the symmetric gauge ${\bf A} = \frac12 {\bf B} \times {\bf r}$
the total angular momentum projection $M_z$, an eigenvalue
of $\hat{L}_z=\sum_i ({\bf r}_i \times -i\hbar\bbox{\nabla }_i)_z$,
is an exact quantum number. In a uniform ${\bf B}=(0,0,B)$
the Hamiltonian (\ref{H}) is also invariant under
a group of magnetic translations whose generators are the components
of the operator $\hat{\bf K} = \sum_{j} \hat{\bf K}_j$,
$\hat{\bf K}_j =
\hat{\bbox{ \pi }}_j - \frac{e_j}{c} {\bf r}_j \times {\bf B}$
(see, e.g., Refs.~\cite{G&D,Simon}).
The operator
$\hat{\bf K}$ is an exact integral of the motion, $[H, \hat{\bf K}]=0$,
whose components commute in $B$ as
\begin{equation}
        \label{comK}
 [\hat{K}_x, \hat{K}_y] = - i \frac{\hbar B}{c} Q \quad , \quad
 \quad Q \equiv \sum_j e_j  \, ,
\end{equation}
while $[\hat{ K}_{ip},\hat{\pi}_{jq}]=0$, $p,q=x,y$.
For neutral complexes (atoms, excitons, biexcitons) $Q=0$
and classification of states in ${\bf B}$ is due to the two-component
continuous vector --- the 2D magnetic momentum
${\bf K}= (K_x,K_y)$ \cite{G&D,Simon}.
For charged systems $Q  \ne 0 $ and the components
of $\hat{\bf K}$ do not commute, which determines the macroscopic
Landau degeneracy of exact eigenstates of (\ref{H}).
Using a dimensionless operator
$\hat{{\bf k}} = \sqrt{c/\hbar B |Q|} \, \hat{\bf K}$
whose components are canonically conjugate,
one obtains raising and lowering Bose ladder operators
$\hat{k}_{\pm}= (\hat{k}_x  \pm i \hat{k}_y)/\sqrt{2}$:
$[\hat{k}_{-}, \hat{k}_{+}]=Q/|Q|$
(see (\ref{dghtr}) below). Therefore,
$\hat{{\bf k}}^2 = \hat{k}_{+} \hat{k}_{-} + \hat{k}_{-} \hat{k}_{+}$
has the oscillator eigenvalues $2k+1$, $k=0, 1, \ldots$.
Since $[\hat{{\bf k}}^2, H]=0$ and $[\hat{{\bf k}}^2, \hat{L}_z]=0$,
the exact charged eigenstates of (\ref{H}) can be simultaneously labeled
by the discrete quantum numbers $k$ and $M_z$ \cite{Simon}.

     The usual optical selection rules for the dipole-allowed transitions
in the Faraday geometry (the light propagates along ${\bf B}$)
are conservation of spin and $\Delta M_z = \pm 1$
for left- and right-circularly polarized light $\sigma^{\pm}$.
There is an additional selection rule: the quantum number $k$ is conserved.
Indeed, the Hamiltonian
$\hat{V}^{\pm} = \sum_{i}
(e_i {\cal F}_0 \hat{ \pi }_i^{\pm}/ m_i \omega) e^{-i \omega t}$
describing the interaction with the light of polarization $\sigma^{\pm}$
(${\cal F}_0$ is the radiation electric field,
$\hat{\pi}_i^{\pm} = \hat{\pi}_{ix} \pm i \hat{\pi}_{iy}$)
commutes with $\hat{\bf K}_i$ and, therefore
\begin{equation}
        \label{comV}
[\hat{V}^{\pm} , \hat{\bf k}^2 ]=0 \:\:\: \Rightarrow \:\:\:
k \:\:\: \mbox{\rm is conserved}\, .
\end{equation}
In fact the perturbation $\hat{V}=F(\hat{\bbox{ \pi }}_i,t)$ can be an
arbitrary function of the kinematic momentum operators
$\hat{\bbox{ \pi }}_i$ and time $t$.
In some limiting cases (e.g., at low fields $B$) $k$ can be associated
\cite{Simon} with the center of the cyclotron motion of a charged system
as a whole. This gives some physical insight into its conservation.
This selection rule is applicable to any charged $e$--$h$ system in $B$.
In particular, it applies to mobile complexes --- charged excitons
$X^-$, $X^+$, multiply-charged excitons $X^{-n}$ (i.e., bound complexes
$(n+1)e$--$h$ with $n>1$, which can exist in special quasi-2D geometries
\cite{Yudson}), and to charged multiple-excitons $X^{-}_{N}$ ($X_N$--$e$),
which exist in 2D systems in high $B$ \cite{AHM}.
In deriving (\ref{comV}) we only used translational invariance
in the plane perpendicular to ${\bf B}$. Therefore,  relation (\ref{comV})
holds in arbitrary magnetic fields and for systems of
different dimensionality, including semiconductors
with a complex valence band described by the Luttinger Hamiltonian \cite{Lutt}.
The selection rule (\ref{comV}) is also valid for internal transitions in
electron systems. Note that for a translationally invariant one-component
(e.g., electron) system with constant charge-to-mass ratio
$e_i/m_i={\rm const}$,
the well-known Kohn theorem \cite{Kohn} states that internal transitions
can occur only at the bare electron cyclotron ($e$--CR) energy
$\hbar\omega_{\rm ce}= \hbar eB/m_e c$. This is a consequence
of the operator algebra
$[H,\hat{V}^{\pm}] = \pm \hbar\omega_{\rm ce}\hat{V}^{\pm}$
involving the center-of-mass {\em inter\/}-LL ladder operators.
On the other hand, relation (\ref{comV}) is based on the algebra of
{\em intra\/}-LL ladder operators. However, for one-component systems
the center-of-mass decouples from internal degrees of freedom in $B$
and both theorems --- though based on different operator algebras ---
give  equivalent predictions in this case.

 To make further discussion concrete, we shall
consider transitions in the $\sigma^-$ polarization in a 2D
three-particle $2e$--$h$ system with the Coulomb interactions,
for a simple valence band, and in the limit of high magnetic fields:
\begin{equation}
        \label{highB}
\hbar\omega_{\rm ce}, \hbar\omega_{\rm ch},
|\hbar\omega_{\rm ce} - \hbar\omega_{\rm ch}| \gg
E_0 = \sqrt{\frac{\pi}{2}} \, \frac{e^2}{\epsilon l_B} \, ,
\end{equation}
where $l_B = \left( \frac{ \hbar c }{e B}\right)^{1/2}$.
Then mixing between different LL's can be neglected
and the $X^-$ states can be classified according to
the total electron and hole LL numbers, $(N_eN_h)$.
The corresponding basis for $X^-$ is of the form \cite{Dz_PLA}
$\phi^{(e)}_{n_1 m_1}({\bf r}) \,
 \phi^{(e)}_{n_2 m_2}({\bf R}) \,
 \phi^{(h)}_{N_{h}M_{h}}({\bf r}_{h})$,
and includes different three-particle $2e$--$h$ states
such that the total angular momentum projection
$M_z= N_e - N_h -m_1 -m_2 + M_h$, and LL's
$N_e=n_1+n_2$, $N_h$ are fixed.
Here $\phi^{(e,h)}_{n m}$ are the $e$- and $h$- single-particle
factored wave functions in $B$; $n$ is the LL quantum number
and $m$ is the oscillator quantum number [$m_{ze(h)}= {+ \atop (-)}(n-m)$].
We use the electron relative and center-of-mass coordinates:
${\bf r} = ({\bf r}_{e1} - {\bf r}_{e2})/\sqrt{2}$ and
${\bf R} = ({\bf r}_{e1} + {\bf r}_{e2})/\sqrt{2}$.
Permutational symmetry requires that for electrons
in the spin-singlet $s$ (triplet $t$) state
the relative motion angular momentum $n_1-m_1$ should be even (odd).
To make this basis compatible with magnetic translations,
i.e., to fix $k$, an additional Bogoliubov canonical transformation
should be performed \cite{P99}.

\begin{figure}[!t]
\epsfxsize=2.9in
\begin{center}
\epsffile{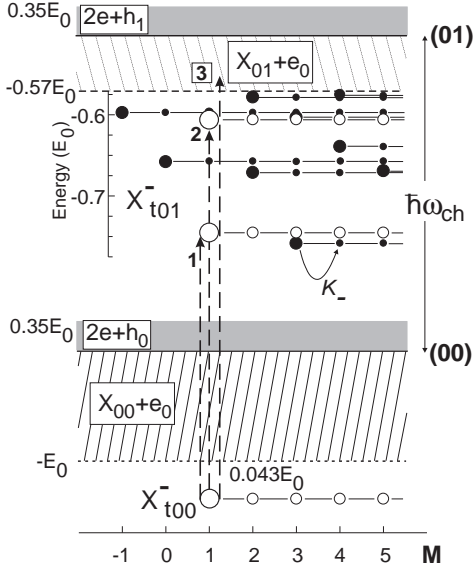}
\end{center}
\caption{
Schematic drawing of bound and scattering electron
triplet $2e$--$h$ states in the lowest LL's ($N_eN_h$)=(00), (01).
The quantum number $M=-M_z$ for the ($N_eN_h$)=(00) states
and $M=-M_z-1$ for the ($N_eN_h$)=(01) states. Large (small) dots
correspond to the bound parent $k=0$ (daughter $k=1,2,\ldots$) $X^-$ states.
Allowed strong transitions in the $\sigma^-$ polarization must satisfy
$\Delta N_h = 1$, $\Delta M_z = -1$, and $\Delta k = 0$. Filled dots in the
(01) LL correspond to families of dark $X^-_{t01}$ states (see text).
}                \label{fig1}
\end{figure}

    The calculated eigenspectra of the three-particle $2e$--$h$ states
with two electrons in the spin-triplet state $t$ ($S_e=1$)
are shown for two lowest ($N_eN_h$)=(00), (01) LL's
in Fig.\,\ref{fig1}.
The spectra consist of continua, which correspond to the motion of
a neutral magnetoexciton (MX) \cite{EP2DS,L&L80}.
In addition, there are discrete bound $X^-$ states lying outside
the continua in which the internal motions of all particles are finite.
The continuum in the ($N_eN_h$)=(00) LL consists of a MX band of
width $E_0$ extending down in energy from the free $(00)$
LL and corresponding to the $1s$ MX ($N_e=N_h=0$) \cite{L&L80}
plus a scattered electron in the zero LL, labeled $X_{00}+e_0$.
In the next hole LL ($N_eN_h$)=(01),
the MX band is of width $0.574 E_0$ \cite{L&L80} and corresponds
to the $2p^-$ MX ($N_e=0$, $N_h=1$)
plus a scattered electron in $N_e=0$ LL, labeled $X_{01}+e_0$.
Moreover, there is a band above each free LL originating from the bound
internal motion of two 2D electrons in $B$
(labeled $2e+h_{N_h}$) \cite{EP2DS}.
The spectra of the discrete bound $X^-$ states are the following.
In the lowest ($N_eN_h$)=(00) LL there exists only one bound
$X^-_{t00}$ triplet state, lying below the lower edge of the $1s$ MX band,
the binding energy of which is $0.043 E_0$ \cite{AHM,Whit97}.
In the next hole LL ($N_eN_h$)=(01) the $2p^-$ MX band is narrow, and
there appear {\it many} bound $X^-_{t01}$ states lying below the continuum
edge (Fig.\,\ref{fig1}). This should be contrasted
with the situation in the next electron LL ($N_eN_h$)=(10),
where only one bound triplet state $X^-_{t10}$ exists \cite{EP2DS}.

In the $\sigma^-$ polarization, the $h$--CR--like inter-LL
$\Delta N_h = 1$ transitions are strong and gain strength in $B$.
These are allowed by the usual selection rules:
spin conserved, $\Delta M_z= -1$.
Consider first the photoionizing $X^-$ transitions in which
the final three-particle states belong to the $X_{01}+e_0$ continuum.
There is an onset at the edge indicated in Fig.\,1
by transition~3. It occurs at an energy
$\hbar\omega_{\rm ch} + 0.469 E_0$, i.e. above the $h$--CR at
an energy that equals the difference in the $1s$ and $2p^-$
MX binding energies, plus the $X^-_{t00}$ binding energy.
This transition may be thought of as the $1s\rightarrow 2p^-$
internal transition of the MX \cite{Dz97},
which is shifted and broadened by the presence of the second electron.
Photoionizing transitions to the $2e+h_1$ band have extremely
small oscillator strengths and can be neglected.

    In order to understand bound-to-bound transitions
$X^-_{t00} \rightarrow X^-_{t01}$, let us describe the structure
of the $X^-$ states in more detail. Generally, there exist
{\em families} of macroscopically degenerate $X^-$ states in $B$.
Each $i$-th family starts with its {\it Parent State} (PS)
$|\Psi^{(P_i)}_{M_z}\rangle$, which has $k=0$ and $M_z$ has its
maximal possible value for that family (cf.\ with translationally invariant
states in 2D electron systems in strong $B$ \cite{Kivelson}).
The normalized daughter states with $k=1, 2, \ldots $ in the $i$-th family,
$|\Psi^{(D_i)}_{M_z'}\rangle$, are constructed
from the PS with the help of the ladder operators:
\begin{equation}
        \label{dghtr}
 | \Psi^{(D_i)}_{M_z - l} \rangle =
\frac{1 }{ \sqrt{l!} }  \hat{k}_{-}^l
      |\Psi^{(P_i)}_{M_z} \rangle  \, ,
\end{equation}
where we have used $[\hat{L}_{z} , \hat{k}_{-}] = -\hat{k}_{-} $.
Conservation of $k$ implies therefore that internal transitions
in the $\sigma^{\pm}$ polarization,
satisfying the usual selection rules $\Delta M_z =\pm 1$, spin conserved,
are allowed only between states from families such that their PS's
are connected by a dipole transition
$|\Psi^{(P_j)}_{M_z}\rangle \rightarrow |\Psi^{(P_i)}_{M_z'}\rangle $,
i.e., have $M_{z}' = M_{z} \pm 1$.
Indeed, for the transition dipole matrix element between the daughter
states in the $m$-th and $n$-th generations we have from (\ref{dghtr}):
\begin{eqnarray}
        \label{D}
{\cal D}_{ij} & = & \langle \Psi^{(D_i)}_{M_z' - m } |
           \hat{V}^{\pm} |\Psi^{(D_j)}_{M_z- n} \rangle =  \\
  & & \frac{1}{\sqrt{n!m!}}   \nonumber
   \langle \Psi^{(P_i)}_{M_z'} | \hat{k}_+^m \hat{V}^{\pm} \hat{k}_{-}^n
           |\Psi^{(P_j)}_{M_z} \rangle \, .
\end{eqnarray}
From the commutativity
$[\hat{V}^{\pm} , \hat{k}_{-} ] = [\hat{V}^{\pm} , \hat{k}_{+} ] =0 $
we see that either $\hat{k}_{-}$ annihilates the left-PS ($n>m$)
or $\hat{k}_{+}$ annihilates the right-PS ($n<m$) in (\ref{D}).
Therefore,  ${\cal D}_{ij}=0$, unless $n=m$ and $M_z' - M_z = \pm 1$.
From the operator algebra it is clear that ${\cal D}_{ij}$ is the same
in all generations and, thus it characterizes the two families of states.
\begin{figure}[!t]
\epsfxsize=3.3in
\begin{center}
\epsffile{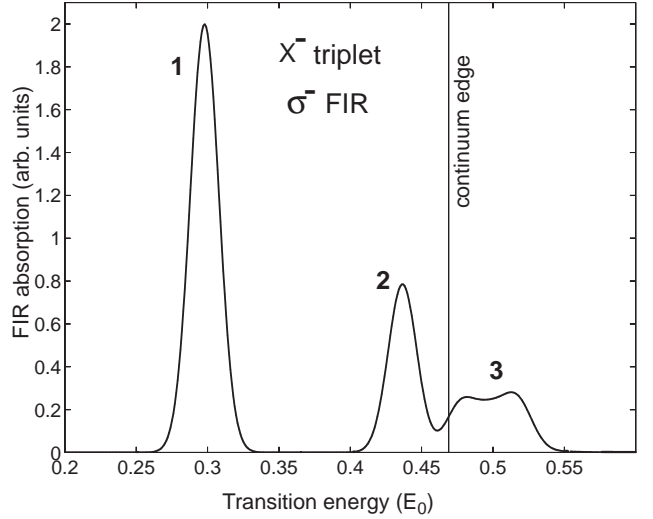}
\end{center}
\caption{
Energies (counted from $\hbar\omega_{\rm ch}$, in units
of $E_0= \protect\sqrt{\pi/2} \, e^2/\epsilon l_B$)
and dipole matrix elements of the inter-LL $\Delta N_h=1$ transitions
(Fig.\ 1) from the ground triplet $X^-_{t00}$ state in the high-field limit.
The spectra have been convoluted with a Gaussian of the $0.02E_0$ width.
}                \label{fig2}
\end{figure}

  This selection rule, due to the rich structure of the continuum,
is easily satisfied for bound-to-continuum transitions.
However, considering many families of the
bound $X^-_{t01}$ states in the next hole LL ($N_eN_h$)=(01),
we see that there are only two PS's that are connected by the
FIR transition (Figs.\,\ref{fig1},\ref{fig2}).
Therefore all families except these two are {\em dark\/},
i.e., are not accessible by internal transitions from
the ground $X^-_{t00}$ bound states:
There exist only two strong bound-to-bound
$X^-_{t00} \rightarrow X^-_{t01}$
transitions in the $\sigma^-$ polarization,
both lying above the $h$--CR (cf.\ \cite{EP2DS}).
Breaking of translational invariance (e.g., by impurities)
would make transitions to many dark states possible
and lead to drastic changes in the absorption spectra.
We note also that in 2D systems
the $X^-$ states belonging to higher LL's are genuinely discrete
only in sufficiently strong $B$ [cf.\ (\ref{highB})]. With decreasing $B$,
the discrete states merge with the MX continuum
of lower LL's and become resonances. Such a situation is also
typical for bulk 3D systems, where $X^-$ (and $H^-$) states in higher
LL's always merge with the continuum of the unbound
internal $z$-motion in lower LL's \cite{Simon}.
Since the quantum number $k$ is still rigorously conserved,
the selection rule predicts which of the resonances are
dark and which are not. The absorption spectra in such situation
can have an asymmetric form typical for Fano-resonances.

In conclusion, we have demonstrated that internal optical transitions
of charged mobile complexes and ions in magnetic fields
are governed by a novel exact selection rule, a manifestation
of magnetic translational invariance. Internal
bound-to-bound transitions of charged excitons $X^-$
in 2D systems in $B$ should be very sensitive to breaking of
translational invariance (by impurities, disorder etc.).
This can be used for studying the extent of $X^-$ localization
in quantum wells.

We thank S.M.\ Apenko, B.D.\ McCombe, D.M.\ Whittaker, and D.R.\ Yakovlev
for helpful discussions.
This work was supported by RBRF Grant \#97-2-17600 and
the ``Nanostructures'' Grant \#97-1072.
ABD is grateful to the Humboldt Foundation for research support.

\end{document}